# Acoustic Scene Classification Using Bilinear Pooling on Time-liked and Frequency-liked Convolution Neural Network

Xing Yong Kek
*Faculty of Science, Agriculture & Engineering*
*Newcastle University*
Singapore
x.y.kek2@newcastle.ac.uk

Cheng Siong Chin
*Faculty of Science, Agriculture & Engineering*
*Newcastle University*
Singapore
cheng.chin@ncl.ac.uk

Ye Li
*Xylem Inc*
Singapore
ye.li@xyleminc.com

*Abstract*— The current methodology in tackling Acoustic Scene Classification (ASC) task can be described in two steps, preprocessing of the audio waveform into log-mel spectrogram and then using it as the input representation for Convolutional Neural Network (CNN) [1]. This paradigm shift occurs after DCASE 2016 where this framework model achieves the state-of-the-art result in ASC tasks [2] on the (ESC-50) dataset and achieved an accuracy of 64.5%, which constitute to 20.5% improvement over the baseline model, and [3] DCASE 2016 dataset with an accuracy of 90.0% (development) and 86.2% (evaluation), which constitute a 6.4% and 9% improvements with respect to the baseline system. In this paper, we explored the use of harmonic and percussive source separation (HPSS) [4] to split the audio into harmonic audio and percussive audio, which has received popularity in the field of music information retrieval (MIR). Although works have been done in using HPSS as input representation for CNN model in ASC task [5], [6], this paper further investigate the possibility on leveraging the separated harmonic component and percussive component by curating 2 CNNs which tries to understand harmonic audio and percussive audio in their 'natural form', one specialized in extracting deep features in time biased domain and another specialized in extracting deep features in frequency biased domain, respectively. The deep features extracted from these 2 CNNs will then be combined using bilinear pooling [7]. Hence, presenting a 'two-stream' time and frequency CNN architecture approach in classifying acoustic scene. The model is being evaluated on DCASE 2019 sub task 1a dataset and scored an average of 65% on development dataset, Kaggle Leadership Private and Public board.

*Keywords—acoustic scene classification; convolution neural network; deep learning; source separation; bilinear pooling.*

## I. INTRODUCTION

Over the past few years, there has been an increase in interest in the research on acoustic/audio analysis [8]. The possible reasons for this phenomenon can be linked to the break-through from visual classification [1] and more publicly available platform and communities for researchers or practitioner to build and benchmark their classification models and discuss their findings and architecture. Detection and Classification of Acoustic Scenes and Events (DCASE) [9] is one such platform, and acoustic scene classification has been one of the on-going challenges that appear in DCASE challenges throughout the years.

Acoustic Scene Classification (ASC) is in general, the ability to computationally differentiate various scene (e.g. beach, airport, streets, shopping centre and etc.) from a given audio signal and the potential of such research is vast (e.g. audio surveillance systems [10], robots or autonomous mobile [11] and environmental noise monitoring [12], [13]).

However, ASC is a challenging task, an audio recording from natural environment can contain multiple sound sources, ambient sounds and each sound sources can also have differing characteristics [10]. This sound sources are usually interleaving and overlapping in a single audio recording, proofing it difficult in identifying the number of sound sources and even if we can separate the sound source, recognizing the quasi-stationary and non-stationary sounds is also challenging due to their changing nature.

A popular paradigm in tackling ASC tasks can be broken down into two steps. First step which is pre-processing of the audio signal and second steps, using the feature extracted from preprocessing fit it into a classifier. The sentiment of DCASE 2018 ASC task 1 challenge has shown favor in the use of log mel-scaled spectrogram as the pre-processing technique and Convolution Neural Network (CNN) as the classifier and has shown promising results with top models achieving about 80% accuracy on DCASE 2018 Task 1. Hence, by following the leads from DCASE 2018, we created our "novel" model which address the important of both temporal information and frequency by using HPSS to best isolate harmonic and percussive component of an audio signal and further perform curated deep feature extraction using 'Time' specified CNN on the harmonic component and 'Frequency' specified CNN on the percussive component and finally combining this 2 features together using bilinear pooling. We present a 'two-stream' architecture for sound classification. The remainder of this paper will be organized as such, Section II describe related works that inspire us to build our model, Section III elaborate on the feature extraction process, Section IV provide details on the model architecture, Section V explained about the experimental setup and dataset used. Section VI give a summary of the test results and Section VII conclude this paper.

## II. MODEL DESIGN

In this paper, we present the use of harmonic percussive source separation (HPSS) [14] as the feature extraction technique and employing a CNN model on harmonic feature and a different CNN model on the percussive feature. Lastly, combining the two CNN models through bilinear pooling [7]. The choice of HPSS was inspired by the work of [5] and [6] who explored the used of HPSS for ASC tasks in DCASE 2018 and demonstrated the possibility of substituting HPSS over direct application of log-mel scale spectrogram.

Though the application of HPSS by [5] and [6] is the combining of harmonic feature and percussive feature as a 2-channel feature input into a CNN model, this paper intuition is that harmonic feature and percussive feature should be processed by different CNN models with more affinity to the nature of the harmonic and percussive feature.

Hence, this paper explored the concept of a specified harmonic CNN (H-CNN) and percussive specified CNN (P-CNN) based on the idea driven by the work of [11], [14], [15] and [16]. These authors have a strong intuition that the current popular approached seen in DCASE 2018 of using a VGG [17] liked CNN as a classifier is not the best CNN model for sound classification tasks. Their reason behind this belief is that VGG liked CNN only employs a (3 × 3) convolution filter throughout its architecture, hence, with a small (3 × 3) local receptive field, it will impede the capability of the model in capturing the temporal relationship. (e.g., VGG which uses (3 × 3) filters will only capture the temporal relationship of 3 frames).

Of course, the simplest solution in solving this problem is by introducing a large convolution filter; unfortunately by doing so will also increase the computational complexity of the model. (e.g., 3×3 filters will have 9 weights and 7×7 filters will have 49 weights) Hence, these authors explored various filters size and concluded a "Time CNN" to have a long horizontal filter and "Frequency CNN" to have a long vertical filter. In conjunction with their findings, we proposed our H-CNN model to adopt a "Time CNN" filter strategy and P-CNN model to adopt a "Frequency CNN" filter strategy. Lastly, the effectiveness of combing these two models have been proven by [15], giving us the direction in combining these two models using bilinear pooling approach [7].

### III. FEATURE EXTRACTION USING HPSS

This paper focus on exploiting the source separated from HPSS as input representation for the CNN model. HPSS was initially used on music classification where the idea is to separate harmonic sound instrument like, e.g. piano, wind instrument, and violin from a percussive sound instrument like e.g. drum and cymbal.

HPSS via median filtering is being selected as it is the most efficient approached [4]. The process of HPSS [4] in this setup can be broken down into four steps. Firstly, the Short-time Fourier transform (STFT) is applied to the raw audio signal to get a time-frequency representation. Secondly, median filtering with a kernel size of (1 × 31) is applied horizontally across the output from the first step to getting harmonic enhanced filtered spectrogram, and likewise, median filter with the similar configuration of (31 × 1) is also applied vertically to get percussive enhanced filtered spectrogram. Thirdly, with the output from the second step,, a masked binary is constructed, and the masked binary will be applied to the original STFT to get the harmonic component and percussive component. Lastly, both components are being processed into a log-mel spectrogram as shown in Fig. 1.

Unlike the usual HPSS performed on single audio to extract both harmonic and percussive features, this paper proposed a frequency biased HPSS to extract percussive feature and a time biased HPSS to extract harmonic feature. The difference between an HPSS for a percussive feature and the harmonic feature can be seen at the very first step, which is the STFT phase. By exploiting the weakness of STFT in regard to the tradeoff of bigger window size will result in a higher resolution in the frequency domain at the expense of losing some temporal details from the time domain. We proposed a huge window size of 1 sec to get the high resolution of frequency before performing median filtering and applying log-mel spectrogram with 256 filter banks. Thus, we determine this feature extraction processed as a frequency biased HPSS while time biased HPSS is treated in a general approach for the STFT with a window size of 42 ms and later applying log-mel spectrogram with 80 filter banks.

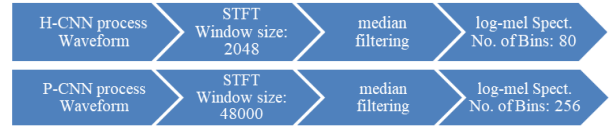

Fig. 1. Block diagram of feature extraction process.

In this experiment, LibROSA [18] is being used as the package for HPSS and converting it to a log-mel spectrogram.

### IV. CNN BILINEAR POOLING ARCHITECTURE

The proposed CNN architecture will consist of 2 CNN models, harmonic CNN, which will be described in *A*. moreover, percussive CNN which will be described in *B*.

These 2 CNN models will then be fused using bilinear pooling [7] and are named as Bilinear CNN (BCNN) by the author. Bilinear pooling is a technique that combines two CNN models by an outer matrix product on the local descriptors [19]:

$$B(HP) = \sum_{s \in S} h_s^T P_s \in \mathbb{R}^{c \times c} \quad (1)$$

Equation (1) is a $(c \times c)$ vector, where $H = \{h_1, h_2, ..., h_s \in \mathbb{R}^c\}$ and likewise to $P = \{p_1, p_2, ..., p_S \in \mathbb{R}^c\}$ are each a set of local descriptors consisting of rows and columns, and S is the number of feature maps extracted through convolution. Local descriptors $h_S, p_S$ are deep features extracted from H-CNN and P-CNN, respectively. Both CNN features are created from a standard convolution layers with a typical non-linear activation function Rectified Linear Unit (ReLU) [2], [1], [20] and usually follow by a widely used max pooling layer [20] to progressively reduce the spatial dimensionality in an effort to reduce computational power and also to reduce overfitting.

Hence, the output of H-CNN and P-CNN after the convolution layer is formulated as:

$$f_{conv} = I \otimes K_q = \sum_i \sum_j I(i,j).K(i-n, j-m) \quad (2)$$

Where *I* is either the input spectrogram or the feature maps with height of *i* and length of *j* and *K* is the filter which both will perform a matrix dot product while *K* convolute over *I*, denoted by $\otimes$, resulting in the output $f_{conv}$. *K* is a 2-dimensional matrix $(n, m)$ where *n* is the height of the filter and *m* is the length of the filter and $Q = \{q_1, q_2, ..., q_p\}$ is a set of $(n, m)$

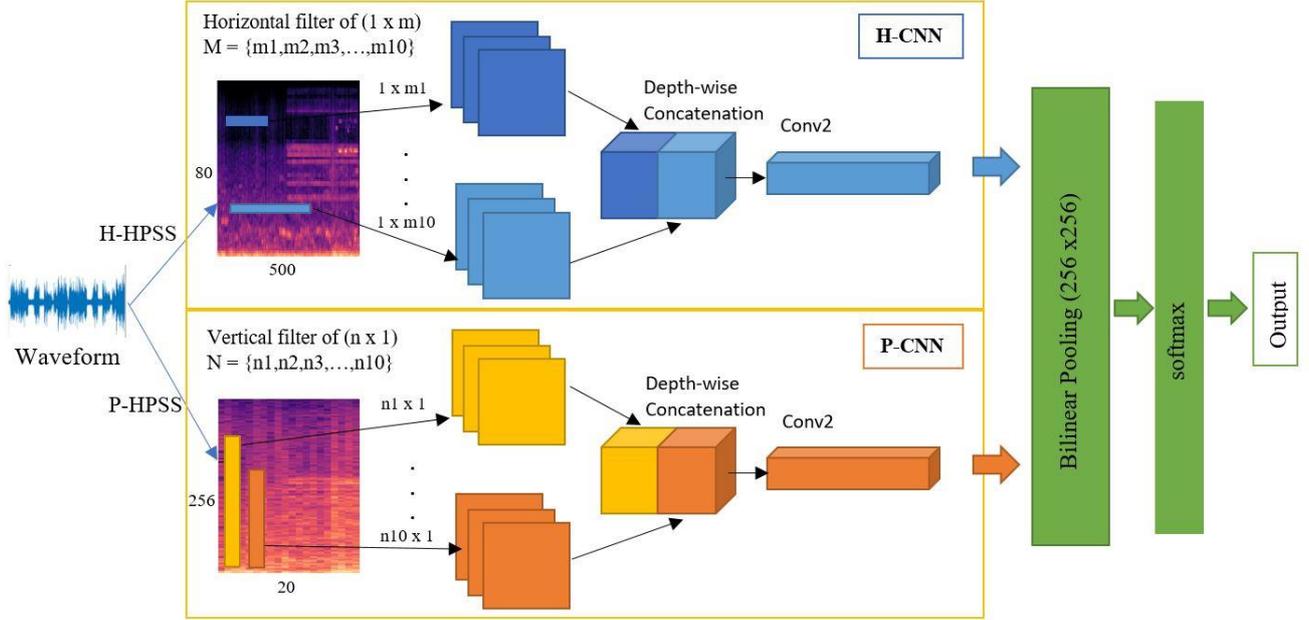

Fig. 2. Proposed CNN bilinear pooling architecture *(only weighted convolution layer is being illustrated)*

Follow by batch normalization (BN) introduced in [21] and applied on ASC task in [3] to tackle internal covariate shift cause by the increase variation between the weight parameters which will result in slower convergence time and possibility of vanishing or exploding gradient. Batch Normalization even out the output from (1) and later an activation function ReLU, $\sigma(x) = \max(0, x)$ [1], [20] is applied, where $x$ is all the features from the output of $f_{conv}$ giving us the below equation:

$$f_r = \sigma(batchNorm(f_{conv})) \quad (3)$$

Lastly, max pooling which also have a filter *KM* with a dimension of $(n, m)$ will convolute over the output $f_r$ while getting the maximum value of the features:

$$f_m = f_r \otimes KM = \sum_i \sum_j \max(f_r(i,j), KM(i-n, j-m)) \quad (4)$$

By combining the building blocks together, we get the output of $h_S$ and $p_S$ as depicted in the pseudo code in Table I and Table II, respectively and the pseudo code of the proposed '2-stream' architecture in Table III.

However, unlike a typical CNN architecture, instead of ending the convolution layers with a fully connected layer or max pooling, this paper proposed the use of bilinear pooling (1) with the intuition to capture pairwise correlations and features interaction, hence increasing the representative power of the deep features [7]. The proposed '2-stream' architecture approach in tackling ASC is illustrated in Fig. 2.

### A. H-CNN (Harmonic CNN) Architecture

Similar to [11] and [15] our model will adopt the concept of Inception module [22] by employing a total of 10 horizontal filters of $(1 \times m)$, hence $Q$ has a size of 10 and each set have a dimension of $(1 \times m_q)$ in (2), which will convolute around the input layer and later combining the features by depth-wise filter concatenation which is illustrated in Fig. 3 and depicted in Table I.

TABLE I. H-CNN PSEUDO CODE

| Pseudo code of H-CNN | |
|---|---|
| **Input:** | Harmonic component log-mel scale spectrogram of dimension (80 x 500 x 1), Filters: $Q = \{(1 \times m_1), (1 \times m_q), ..., (1 \times m_{10}) \in (1 \times \mathbb{Z})\}$ |
| **Output:** | deep features $h_S$ |
| **Conv1:** | foreach filters size in $M$ do     Convolution: $f_{conv} = I \otimes K_q$     BN and ReLU refer to (3) end for Depth-wise concatenation of all the features: $f_d = depthConcat(M \text{ number of } f_r)$ Max pooling: $f_m = f_d \otimes KM$ and $(n,m) = (10,4)$ |
| **Conv2:** | Convolution: $f_{conv} = f_m \otimes K$ and $q = (7,7)$ BN and ReLU Max pooling: $f_m = f_d \otimes KM$ and $(n,m) = (2,2)$ |
| **Output:** | $h_S = f_m$ |

TABLE II. P-CNN PSEUDO CODE

| Pseudo code of P-CNN | |
|---|---|
| **Input:** | Harmonic component log-mel scale spectrogram of dimension (256 x 20 x 1), Filters: $Q = \{(1 \times n_1), (1 \times n_q), ..., (1 \times n_{10}) \in (1 \times \mathbb{Z})\}$ |
| **Output:** | deep features $p_S$ |
| **Conv1:** | foreach filter $Q$ do     Convolution: $f_{conv} = I \otimes K_q$     BN and ReLU end for Depth-wise concatenation of all the features: $f_d = depthConcat(Q \text{ number of } f_r)$ |
| **Conv2:** | Convolution: $f_{conv} = f_m \otimes K$ and $q = (5,5)$ BN and ReLU Max pooling: $f_m = f_d \otimes KM$ and $(n,m) = (10,2)$ |
| **Output:** | $p_S = f_m$ |

The intuition is to capture temporal relationship using various large horizontal filters from the harmonic source of the acoustic scene. The H-CNN model configuration is detailed in Table IV.

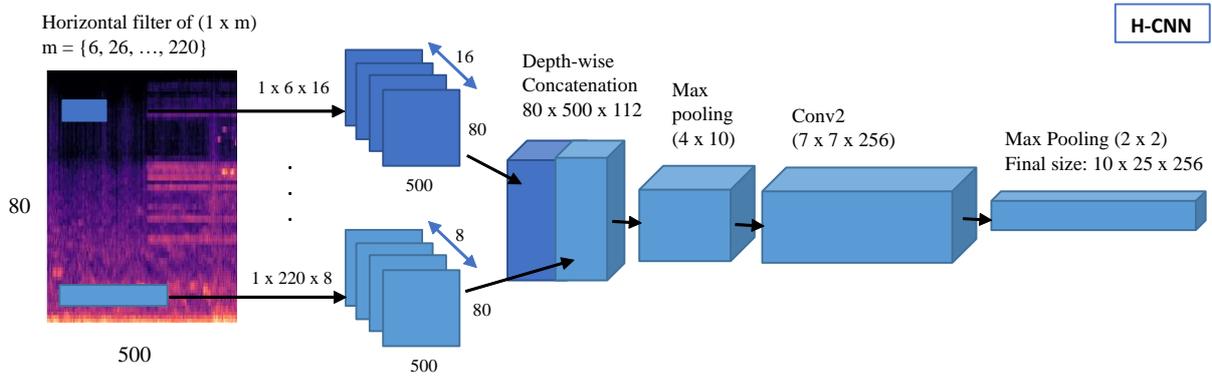

Fig. 3. H-CNN model architecture *(only 2 horizontal filters is being illustrated in this diagram)*

TABLE III. PROPOSED CNN BILINEAR POOLING PSEUDO CODE

| Pseudo code of proposed CNN bilinear pooling | |
|---|---|
| **Input:** | Harmonic features: $h_S$ <br> Percussive features: $p_S$ |
| **Output:** | Result of the weighted score of 10 scenes |
| **Bilinear Pooling** | Bilinear Pooling: $B(HP) =$ <br> $Reshape(h_S \& p_S)$ from 3-dimension to 2-dimension by combining the ($height \times length$): output H & P with ($hl \times c$) where $c$ is the number of channels/depths and $hl = (height \times length)$ <br> $B(HP) = H(hl \times c).P(hl \times c)^T$ <br> signed square-root(B) <br> $\ell_2$normalization(B) |
| **Output:** | Result = softmax($B$) a multiclass logistic classifier [20] |

### B. P-CNN (Percussive CNN) Architecture

Likewise, for P-CNN, the concept of Inception module is adopted, unlike H-CNN which specialized in capturing temporal relationship, P-CNN employs 10 vertical filters of $(n \times 1)$, hence $Q$ has a size of 10 and each set have a dimension of $(n_q \times 1)$ in (2), where n denotes the height of each filter. The idea here is to capture frequency which is the pitch aspect of the acoustic scene and the configuration of the CNN model is detailed in Table V.

## V. EXPERIMENTAL SETUP

This paper uses dataset from DCASE 2019 Task 1a [23] which topic is on acoustic scene classification and researchers are given a dataset containing audio files to perform training and classification on 10 acoustic scenes. The 10 acoustic scenes being the airport, shopping mall, metro station, street pedestrian, public square, street traffic, tram, bus, metro and park.

### A. Dataset

The development dataset for this task is the TAU Urban Acoustic Scenes 2019 Development dataset**,** is a collection of 10s audio clips which is recorded from 4 devices with a 48khz sampling rate and 24-bit resolution. The recording is conducted in 10 European cities and at 10 different location. A total of 40 hours of audio, with 14400 segments are collected in the development dataset and is evenly balanced between scene classes (144 segments per city per acoustic scene).

The trained model will then be evaluated on another 2 datasets, which is the TAU Urban Acoustic Scenes 2019 Leaderboard dataset and Evaluation dataset with Leaderboard dataset having a 3GB worth of 10s audio clips and Evaluation dataset containing 20 hours of audio recordings.

TABLE IV. A TOP TO BOTTOM PERSPECTIVE OF H-CNN MODEL CONFIGURATION

| Layer | H-CNN filters configuration |
|---|---|
| **Input 1** | 80 x 500 |
| **Conv1** | 16, (1,6) ; 16, (1,26) ; 16, (1,50) ; 16, (1,76) ; 8, (1,96) |
| | 8, (1,120) ; 8, (1,146) ; 8, (1,170) ; 8, (1,196) ; 8, (1,220) |
| **Filter Concatenate** | 80 x 500 x 112 |
| **Max Pooling** | (4,10) with stride of 1 |
| **Conv2** | 256, (7,7) |
| **Max Pooling** | (2,2) with stride of 1 |
| **Output** | 10 x 25 x 256 |

### B. Training Setup and Evalution Setup

The training set consists of 9185 segments and 4185 segments for the testing set, which gives us the setup of splitting the development dataset by ~70% for training the model and ~30% for testing. For this challenge, the full 10s audio clip is being used, and no down sampling is being performed. Each audio clip is first, being converted from stereo into the mono channel and HPSS is being applied to split the audio into the harmonic wave and percussive wave.

Next, we preprocessed both waves into log-mel spectrogram with 50% overlapping, which will be the input representation of the network. The setup for the model training is as such; the model is trained with a mini-batch size of 16 samples over 200 epochs with patience of 15 epochs, meaning the model will stop training if validation loss did not improve after 15 epochs. Adam optimizer is being applied to the model with 0.001 learning rate, and learning rate will start to reduce if validation loss plateau for 5 epochs. The trained model is then being tested on the evaluation dataset.

## VI. RESULT AND DISCUSSION

We compare our model against Dcase2019 baseline model [23] on 2 datasets, Dcase2019 task 1a development test set, and Kaggle Leaderboard dataset depicted in Table VII. and a breakdown of accuracy result on development test set based on scene is depicted in Table VI.

TABLE V. A TOP TO BOTTOM PERSPECTIVE OF H-CNN MODEL CONFIGURATION

| Layer | P-CNN filters configuration |
|---|---|
| Input 1 | 256 x 20 |
| Conv1 | 32, (10,1) ; 32, (25,1) ; 32, (42,1) ; 16, (70,1) ; 16, (100,1) |
| | 16, (110,1) ; 16, (120,1) ; 16, (150,1) ; 16, (160,1) ; 16, (220,1) |
| Filter Concatenate | 256 x 20 x 208 |
| Conv2 | 256, (5,5) |
| Max Pooling | (10,2) with stride of 1 |
| Output | 25 x 10 x 256 |

TABLE VI. COMPARSION RESULTS OF PROPOSED MODEL BY SCENE

| Scene | Accuracy Result based on Percentage | |
|---|---|---|
| | Baseline | Proposed |
| Airport | 48.4% | **54.4%** |
| Bus | 62.3 % | **68.7%** |
| Shopping Mall | 59.4 % | **69.3%** |
| Street Pedestrian | **60.9 %** | 60.8% |
| Street Traffic | **86.7 %** | 86.0% |
| Metro Station | 54.5% | **62.5%** |
| Park | 83.1% | **83.4%** |
| Metro | **65.1%** | 57.3% |
| Public Square | 40.7% | **57.4%** |
| Tram | **64.0%** | 62.6% |

Results show that our model performs slightly better than the DCASE 2019 baseline model and demonstrated that the proposed model is a feasible model for ASC tasks.

In a prior experiment, we have trained H-CNN model and P-CNN model separately with both models ending with a global max pooling follow by a softmax to determine the classification output. The accuracy performance achieved by both models is only between 57% to 59%. It means that both frequency and time features are equally important for our proposed model and experiment. Furthermore, this experiment result has proven that bilinear pooling is effective in combining frequency features and time features.

## VII. CONCLUSION

To conclude, we have demonstrated that our proposed model has much potential and is now in its fetus stage. Hence, more work is required in fine-tuning the CNN architecture for both H-CNN and P-CNN model as currently, the number of filters is empirically decided and model enhancement technique such as dropout, data augmentation and hyper-parameter training can be employed to further improve the model. Another improvement will be on exploring other bilinear pooling techniques. As the downside of employing bilinear pooling is its substantial computational cost which is also stated by [7] and [19].

TABLE VII. COMPARSION RESULTS OF PROPOSED MODEL

| Model | Comparison dataset | |
|---|---|---|
| | *Development test set* | *Kaggle leaderboard (Public/Private)* |
| Baseline | ~62% | ~64.4%/~63% |
| Proposed | ~66% | ~64.7%/~66% |

In the feature extraction spectrum, we can investigate on other possible signal processing technique that establishes a more frequency domain bias and time domain bias.

Lastly, further acoustic classification test can be conducted to evaluate the robustness and capability of this model approach as this model should not be limited to only ASC tasks.